# A procedure for observing rocky exoplanets to maximize the likelihood that atmospheric oxygen will be a biosignature.

Steven J. Desch[1,*], Stephen Kane[2], Carey M. Lisse[3], Cayman T. Unterborn[1], Hilairy E. Hartnett[1,4], Sang-Heon Shim[1]

[1]School of Earth and Space Exploration, Arizona State University, PO Box 871404, Tempe AZ 85287-1404;

[2]Department of Earth Sciences, University of California, Riverside, 900 University Ave., Riverside CA 92521;

[3]Applied Physics Laboratory, Johns Hopkins University, 11100 Johns Hopkins Rd., Laurel, MD 20723;

[4]School of Molecular Sciences, Arizona State University, PO Box 871604, Tempe, AZ 85287-1604.

*Corresponding author: steve.desch@asu.edu, (480) 965-7742.

**Introduction**

The search for life on planets around other stars is one of the grand scientific challenges of the 21st century. The approach being adopted by the astronomical community is to find putative biosignature gases—especially oxygen or methane—in the atmospheres of such planets, through infrared transmission or emission spectroscopy (Domagal-Goldman, Wright et al. 2016). To detect these biogases using transit spectroscopy will require at least the sensitivity of the James Webb Space Telescope (*JWST*), to be launched in 2019. It appears possible to use *JWST* to characterize the atmospheres of many known transiting rocky exoplanets around M stars (Morley et al. 2017), but there are likely to be many more habitable planets to characterize than *JWST* can observe.

The astronomical community therefore must plan these observations carefully, in terms of target selection and interpretation. The Transiting Exoplanet Survey Satellite (*TESS*) will find dozens of transiting exoplanets in their star's habitable zones (Sullivan et al. 2015), and a strategy will be needed to prioritize them for follow-up *JWST* observations. From nearby stellar population statistics and Kepler planet frequency results (Mulders et al. 2015), most (perhaps all) transiting rocky exoplanets likely will be found in the habitable zones of M dwarfs (e.g., Trappist-1e; Gillon et al. 2017), with water contents possibly much greater than Earth. Such planets appear more prone to false positives for oxygen (e.g., Luger & Barnes 2015). Modeling will be needed to interpret whether oxygen is a true biosignature on various exoplanets, and to prioritize observations of exoplanets.

Here we advocate an observational strategy to help prioritize exoplanet observations. It starts with more easily obtained observational data, and ranks exoplanets for more difficult follow-up observations based on the likelihood of avoiding planets for which oxygen is a false positives or even an inconclusive signature of life. We find that for oxygen to be a robust biosignature, both land and surface water must be present. Landless exoplanets have much slower biogeochemical cycles, so while oxygenic photosynthesizing life could exist on such planets, it could not produce oxygen at a rate competitive with abiotic rates such as photolysis. These habitable planets, whose life would not be *detectable*, should be avoided.

**Too much water obscures the signs of life**

Life as we know it requires water, and water is equated with habitability. NASA's mantra in astrobiology has been to "Follow the Water". We recently presented work at the November 2017 Habitable Worlds Meeting demonstrating that too much water is detrimental to oxygen production by life (Desch et al., in preparation). On an Earth-like planet with 50 oceans (~1 wt% bulk $H_2O$), continents and geochemical cycles take place under a thick (~100 km) high-pressure ice mantle, cut off chemically from the oceans; the pressure of 100 oceans (~2 wt% $H_2O$) can suppress silicate melt and stop outgassing and geochemistry altogether (Unterborn & Schaefer, in preparation). Just 5 oceans (~0.1wt% $H_2O$) are enough to submerge the continents and slow the influx of phosphate (by chemical dissolution of felsic rocks) into the oceans (because of the higher pH of the oceans compared to rainwater), by 2-3 orders of magnitude. This limits the biogenic export of oxygen into the atmosphere to levels comparable to or exceeded by rates of abiotic photolysis and hydrogen escape. **To be certain that oxygen in a planet's atmosphere is biological in origin, one must observe a planet that has both water and land on its surface, i.e., one with < 5 oceans' worth of water (< 0.1wt% $H_2O$).**

**An observational procedure for observing exoplanets**

Current techniques can identify planets with > few wt% $H_2O$, but these should be deprioritized for observations because oxygen will not be a reliable biosignature on them. **Although life as we know it requires water, we should actually search for planets with as little water as possible.** Observations should be undertaken in the order laid out in our flowchart (Figure 1). Here we expand on these steps.

Step 1: Find rocky exoplanets in the habitable zones (HZs) of their stars. Surveys (e.g., by *TESS*) will generate dozens of potential planets. Exoplanets with radii > 1.5 $R_E$ should be excluded because they are likely to have thick $H_2$/He atmospheres (Weiss & Marcy 2014). Exoplanets in multi-planet systems (e.g., Trappist-1) may be preferred, as masses can be much better refined by transit timing variations (TTVs), In the next 15 years, these selection criteria will highly favor exoplanets around M dwarfs (Charbonneau 2017).

Step 2: Determine the current X-ray/ultraviolet (XUV) fluxes and infer the past XUV fluxes of host stars. Flux values exist in the ROentgen SATellite (ROSAT) All Sky Survey catalog, and new, detailed measurements, at least down to 0.1 keV, can be obtained using Chandra or X-ray Multi-Mirror (XMM) observatories. Stars with insufficient past activity to strip $H_2$/He atmospheres should be excluded, as these atmospheres will prevent a determination of water content. All the Trappist-1 planets easily could have lost ~$10^{-2}$ $M_E$ of $H_2$/He atmosphere over 8 Gyr assuming Trappist-1's current XUV flux, much greater than the masses of $H_2$/He gas accreted from their nebula, ~ $10^{-3}$ $M_E$ (Unterborn et al., in revision). The XUV flux also should not be consistent with abiotic $O_2$ buildup by hydrogen escape. Luger & Barnes (2015) find that fluxes must exceed a critical limit so that $O_2$ generated by $H_2O$ photolysis escapes along with the hydrogen. The XUV fluxes must exceed both limits.

Step 3: Measure the planetary masses and radii as precisely as possible. Radii determined from precise photometry will probably not be as uncertain as the masses. Masses should be derived from RV measurements, but since most of the host stars are likely to M dwarfs, TTVs likely will be the stronger constraints on planetary mass, which is why multi-planet systems might be preferred.

Step 4: Obtain stellar abundances. Modeling the mass-radius relationships of exoplanets requires determining the likely range of bulk Fe/Mg the planet might have. The planetary Fe/Mg ratio is the stellar Fe/Mg ratio, modified by estimates of mantle stripping and disk processes. We find that variations across the range of observed stellar compositions (Fe/Mg = 0.4 to 1.5) lead to 20% variations in the mass and density. In principle, < 1wt% $H_2O$ abundances could be inferred if the Fe/Mg ratio were constrained to < 5% or 0.02 dex. This precision is unlikely, but the more precisely the composition is determined, the better.

Other stellar abundances (Mg/Si, Na/Si, K, U, Th) also can provide very important geophysical context, as their abundances in the planet set the vigor of mantle convection, and possibly plate tectonics. These could help prioritize or de-prioritize planets for observations, in concert with sophisticated geophysical modeling.

Step 5: Conduct sophisticated modeling to find the probability that the observed mass and radius of the exoplanet is consistent with "no" (< 0.1wt%) water. Part of this modeling must include new equations of state for rock-water alloys at high pressures, and non-Earth-like compositions. As exoplanets are observed and placed in the mass-radius diagram, those lying above the mass-radius curve for waterless planets (i.e., less dense than pure rock/metal planets) must be inferred to have abundant $H_2O$. Those lying furthest below the curve (in sigmas, the observational uncertainties in mass and radius) are most likely to be waterless planets. Those should be prioritized for further observation.

Step 6: Perform low-spectral-resolution optical transmission spectroscopy to determine the transit depth as a function of color. A lack of variation (as for GJ1214b: Kreidberg et al. 2014) would indicate either the lack of an atmosphere or the presence of hazes. Since observation of the surface is demanded (and of course life needs an atmosphere), only those exoplanets with some variation in transit depth with color should be further characterized.

Step 7: Perform high-spectral-resolution infrared transmission spectroscopy. At this step the sought biosignature oxygen could be detected. $H_2O$ vapor also is detectable and should be found, to signal the presence of liquid water on the surface. $CO_2$ is likely to be present and detected, but if $CO_2$ is present at ~ 1 bar levels, this would indicate a breakdown in the carbonate-silicate cycle, suggesting a dearth of bioessential elements, de-prioritizing the planet for further observations.

Step 8: Obtain the optical reflectance light curve. If and only if the previous steps have indicated a planet with relatively little water—an atmosphere with oxygen, and signs of liquid water on the surface—should the an attempt be made to measure the optical reflectance light curve. Reflected light may be used to find evidence of glint, supporting the presence of liquid water on the surface (Williams & Gaidos 2008), but the most important role of light curve analysis is to identify land. Principal component analysis of the time-varying reflected light in various filters has detected land and oceans on Earth, and could identify patches of land, ocean, and vegetation (Cowan et al. 2009; Fujii et al. 2010).

The procedure outlined above allows observers to start with observations of planetary mass and radius and stellar XUV flux and elemental abundances, and then prioritize only the most promising planets for the more difficult, time-consuming observations involving spectroscopy and reflectance light curves. Presumably all planets observed would be habitability; those with surface water and land would be favored, so that oxygen could be the most robust biosignature if detected.

This exercise highlights potentially mutually exclusive selection criteria. For example, HZ exoplanets around M dwarfs are favored for atmospheric measurements, for the likelihood they transit and the transit depths. But optical light curves are more easily obtained for HZ exoplanets around FGK stars, as M star HZs are within the inner working angle of most telescope designs. Elemental abundances of M dwarfs also are difficult to obtain. These and other factors may need to be weighed against each other in mission development and design.

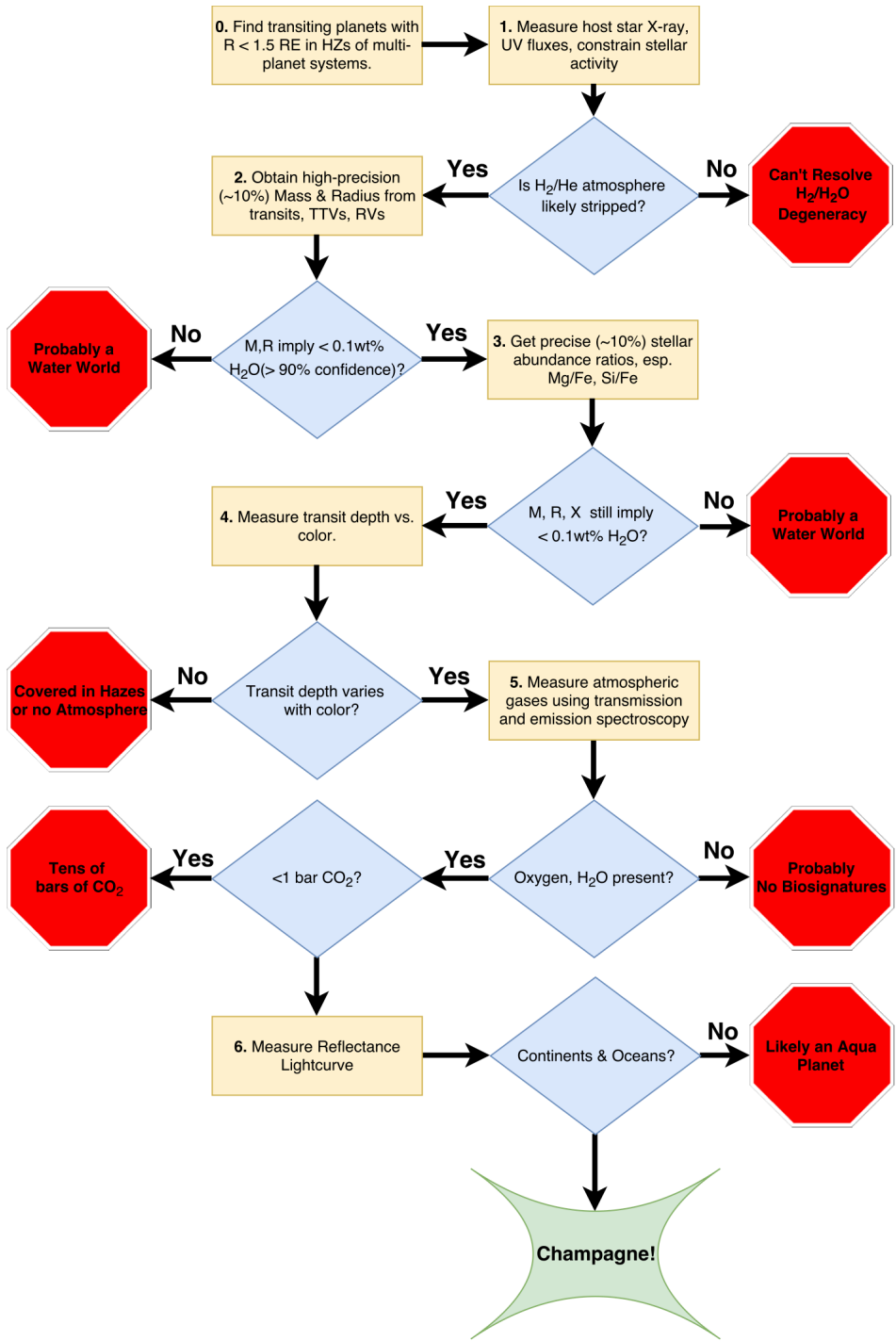

**Figure 1**: A flowchart describing the observational campaign being advocated here, designed to find not just planets with oxygen in their atmospheres, but planets for which oxygen also would be a biosignature. The observations range from those currently being undertaken, to those requiring future space missions and possible for only a handful of exoplanets.